\documentclass[twocolumn, prl, superscriptaddress]{revtex4-1}
\usepackage{amssymb} 
\usepackage{amsmath}
\usepackage{epsfig}
\usepackage{graphics, graphicx}
\usepackage{bbold}
\usepackage{psfrag}
\usepackage{mathcomp}
\usepackage{subfigure}
\usepackage{verbatim}
\usepackage{color}
\usepackage[colorlinks,citecolor=blue]{hyperref}
\def\cp#1{\mathbf{#1}}

\begin{document}

\title{Fermi polaron revisited: polaron-molecule transition and coexistence}
\author{Xiaoling Cui}
\email{xlcui@iphy.ac.cn}
\affiliation{Beijing National Laboratory for Condensed Matter Physics, Institute of Physics, Chinese Academy of Sciences, Beijing, 100190, China}
\affiliation{Songshan Lake Materials Laboratory , Dongguan, Guangdong 523808, China}
\date{August 24, 2020}

\begin{abstract}
We revisit the polaron-molecule transition in three-dimensional(3D) fermion systems  using  the well-established variational approach. The molecule is found to be intrinsically unstable against lowest-order particle-hole excitations, and it can only approximate the ground state of impurity system with finite total momentum in the strong coupling regime.
The polaron-molecule transition can therefore be reinterpreted as a first-order transition between single impurity systems with different total momenta. Within certain interaction window near their transition, both states appear as local minima in the dispersion curve, indicating they can coexist in a realistic system. 
We have further confirmed the polaron-molecule coexistence in the presence of a finite impurity concentration and at low temperature, which directly leads to a smooth polaron-molecule transition as observed in recent experiments of 3D ultracold Fermi gases.
Our results have provided an unambiguous physical picture for the competition and conversion between polaron and molecule, and also shed light on Fermi polaron properties in low dimensions. 
\end{abstract}
\maketitle

{\it Introduction.} Fermi polaron is a quasi-particle describing an impurity dressed by surrounding fermions. In recent years it has attracted great attention and also been successfully realized in the field of ultracold gases\cite{Zwierlein,Salomon,Salomon2,Grimm,Kohl,Grimm2016,Roati,Sagi}, thanks to the high controllability of spin numbers and interaction strength. Nearby a Feshbach resonance, the Fermi polaron exhibits an attractive lower branch \cite{Chevy, Lobo, Combescot1, Combescot2, Prokofev, Leyronas, Punk, Enss, Bruun, Castin, Parish, Parish2} and a repulsive upper branch \cite{Cui, Troyer, Bruun2, Parish3, Demler}. 
These two branches are crucially important for understanding, respectively, the stability of fermion superfluidity and itinerant ferromagnetism in the high polarization limit of fermion system. 

For the attractive Fermi polaron, it is commonly believed that there is a first-order transition between a polaronic state (impurity dressed with fermions surrounded) and a molecular state (impurity bound with one fermion on top of Fermi sea) when the attraction increases, as theoretically studied in both three-dimension(3D)\cite{Prokofev, Leyronas, Punk, Enss, Bruun, Castin} and two-dimension(2D)\cite{Parish,Parish2}. Many theories are based on the energy comparison from two distinct variational  ansatz for polaron and molecule with truncated particle-hole excitations\cite{Leyronas, Punk, Castin, Parish, Parish2}, while an alternative claim of a smooth polaron-molecule crossover was also proposed by arguing that the two ansatz are mutually contained when including more particle-hole excitations\cite{Edwards}. Thus it is important to identify the nature of polaron-molecule transition, if exists, based on a unified treatment for polaronic and molecular states\cite{footnote_MC}. 
Moreover, in the experimental side, previous studies in 3D\cite{Zwierlein} and 2D\cite{Kohl} have reported the polaron-molecule transition with a continuous zero-crossing of quasi-particle residue $Z$, instead of a sudden jump of $Z$ associated with a first-order transition. Furthermore, a very recent experiment of a 3D Fermi gas\cite{Sagi} has observed a smooth evolution of various physical quantities from weak to strong coupling regime, and also pointed to a coexistence between polaron and molecule nearby their transition. As these observations cannot be fully explained by the trap inhomogeneity\cite{Sagi}, the underlying mechanism for the polaron-molecule coexistence is still an open question.  

\begin{figure}[t]
\includegraphics[width=9cm]{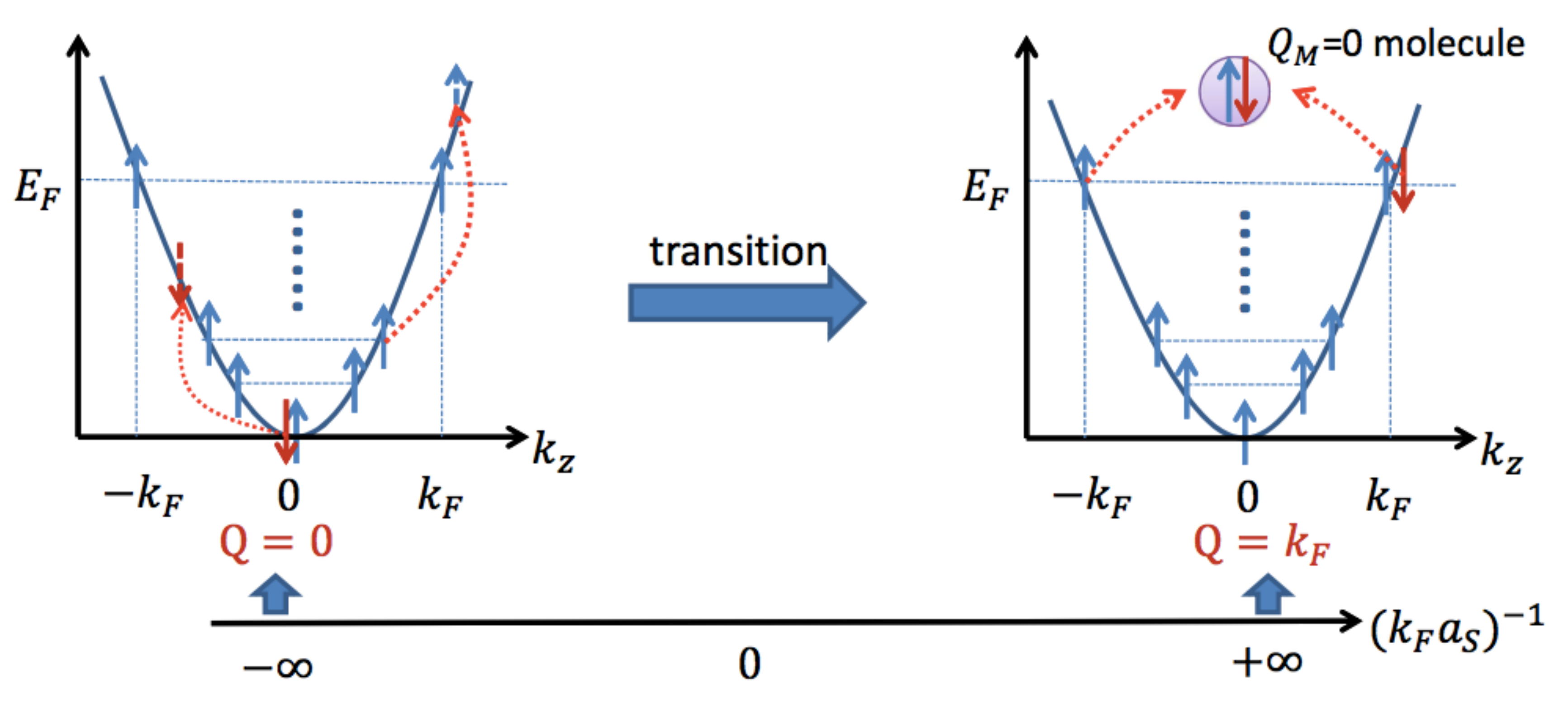}
\caption{(Color Online). Illustration of polaron-molecule transition for a single impurity (spin-$\downarrow$) imbedded in a spin-$\uparrow$ Fermi sea (with Fermi energy $E_F$ and Fermi momentum $k_F$). In the regime $(k_Fa_s)^{-1}\rightarrow-\infty$, the ground state is a zero-momentum polaron dressed by particle-hole excitations. In the regime $(k_Fa_s)^{-1}\rightarrow +\infty$, the ground state switches to total momentum $Q=k_F$, such that the impurity can pair with a spin-$\uparrow$ originally at the Fermi surface to form a deeply bound molecule with zero center-of-mass momentum. Due to the rotational invariance of ${\bf Q}$, here we take it along z-axis and only plot out the $k$-space excitations with $k_x=k_y=0$. }  \label{fig_schematic2}
\end{figure}

To address above issues, in this work we revisit the 3D Fermi polaron problem using the well-established variational approach\cite{Chevy, Combescot1, Combescot2, Leyronas, Punk, Castin, Parish, Parish2, Cui, Parish3}. 
The main contributions are in three fold:

(1) The molecule is found to be intrinsically unstable against lowest-order particle-hole excitations. However, it can serve as a good approximation to describe the impurity system at finite momentum $Q=k_F$ in the strong coupling regime ($k_F$ is the Fermi momentum of majority fermions).
It follows that the literally called polaron-molecule transition can be reinterpreted as a first-order transition between single impurity systems with different total momenta $Q=0$ and $Q=k_F$, see illustration in Fig.\ref{fig_schematic2}. Since different-$Q$ states cannot be smoothly connected by particle-hole excitations, the theoretical debate in Ref.\cite{Edwards} is naturally resolved. 

(2)
Within certain interaction window near their transition, the two $Q$-states are found to appear simultaneously as local minima in the dispersion curve.  This provides the underlying mechanism for polaron-molecule coexistence 
in realistic systems. 

(3)
Taking the realistic condition in experiment with a finite impurity concentration and at low temperature, we have confirmed the polaron-molecule coexistence and reproduced a smooth evolution of all physical quantities as measured in Ref.\cite{Sagi}. This provides an intrinsic reason for the smooth polaron-molecule transition as observed in 3D Fermi gases\cite{Zwierlein,Sagi}, and also sheds light on similar phenomenon in  2D system\cite{Kohl}.

{\it Model.} We consider the following Hamiltonian for the 3D Fermi gases with contact interaction:
\begin{equation}
H=\sum_{\cp k\sigma} \epsilon_{\cp k} c^{\dag}_{\cp k \sigma}c_{\cp k \sigma}+U\sum_{{\cp Q},{\cp k},{\cp k'}} c^{\dag}_{{\cp Q}-{\cp k},\uparrow}c^{\dag}_{{\cp k},\downarrow}c_{{\cp k'},\downarrow}c_{{\cp Q}-{\cp k'},\uparrow} 
\end{equation}
Here $c^{\dag}_{\cp k, \sigma}$ is the creation operator of spin-$\sigma$($\uparrow,\downarrow$) fermion with momentum ${\cp k}$ and energy $\epsilon_{\cp k}={\cp k}^2/(2m)$; $U$ is the bare interaction that can be connected to the s-wave scattering length $a_s$ via $1/U=m/(4\pi a_s)-1/V\sum_{\cp k}m/{\cp k}^2$. For brevity we will take $\hbar=1$ throughout the paper.

For a single $\downarrow$-impurity immersed in the Fermi sea of $\uparrow$-atoms with number $N$ (giving the Fermi momentum $k_F$ and Fermi energy $E_F=k_F^2/(2m))$), 
based on the variational approach\cite{Chevy, Combescot1, Combescot2, Leyronas, Punk, Castin, Parish, Parish2, Cui, Parish3} we can write down a general ansatz for the system with total momentum ${\cp Q}$:
\begin{equation}
|P({\cp Q})\rangle=\left(\phi_0 c^{\dag}_{\cp Q,\downarrow}+\sum'_{{\cp k},{\cp q}}\phi_{{\cp k},{\cp q}} c^{\dag}_{{\cp Q}+{\cp q}-{\cp k},\downarrow}c^{\dag}_{\cp k,\uparrow}c_{\cp q,\uparrow}+...\right) |FS\rangle_{N};  \label{wf_p}
\end{equation}
and the molecule ansatz is:
\begin{equation}
|M(0)\rangle=\left(\sum'_{{\cp k}}\psi_{{\cp k}} c^{\dag}_{-{\cp k},\downarrow}c^{\dag}_{\cp k,\uparrow}+...\right) |FS\rangle_{N-1} \label{wf_m}.
\end{equation}
Here $\sum'$ refers to summation under $|\cp k|>k_F$ and  $|\cp q|\leqslant k_F$; ``$...$" refers to terms with higher-order particle-hole excitations, which are neglected in this work given their destructive interference for the attractive branch\cite{Combescot2}. 

The two ansatz above give rise to very different pictures: (\ref{wf_p}) describes a fermionic quasi-particle where the impurity is dressed by majority fermions with particle-hole excitations, while (\ref{wf_m}) represents a bosonic molecule where the impurity is bound with a single fermion on top of the Fermi sea. 
Given such distinct features and an energy crossing as the attraction is increased, many theories have predicted a first-order transition from $|P(0)\rangle$ to $|M(0)\rangle$\cite{Prokofev, Leyronas, Punk, Enss, Bruun, Castin}, called the {\it polaron-molecule transition}. On the other hand, an alternative claim of a smooth crossover, instead of a sharp transition, was proposed based on the argument that $|M\rangle$ is a special case of $|P\rangle$ family and vice versa if including arbitrarily many particle-hole excitations\cite{Edwards}. 
In this sense, it is highly demanded to re-examine the relation between the two ansatz. Due to the rotational invariance of ${\cp Q}$, from now on we choose a specific case with ${\cp Q}=Q{\cp e}_z$. 

\begin{figure}[t]
\includegraphics[width=7.5cm]{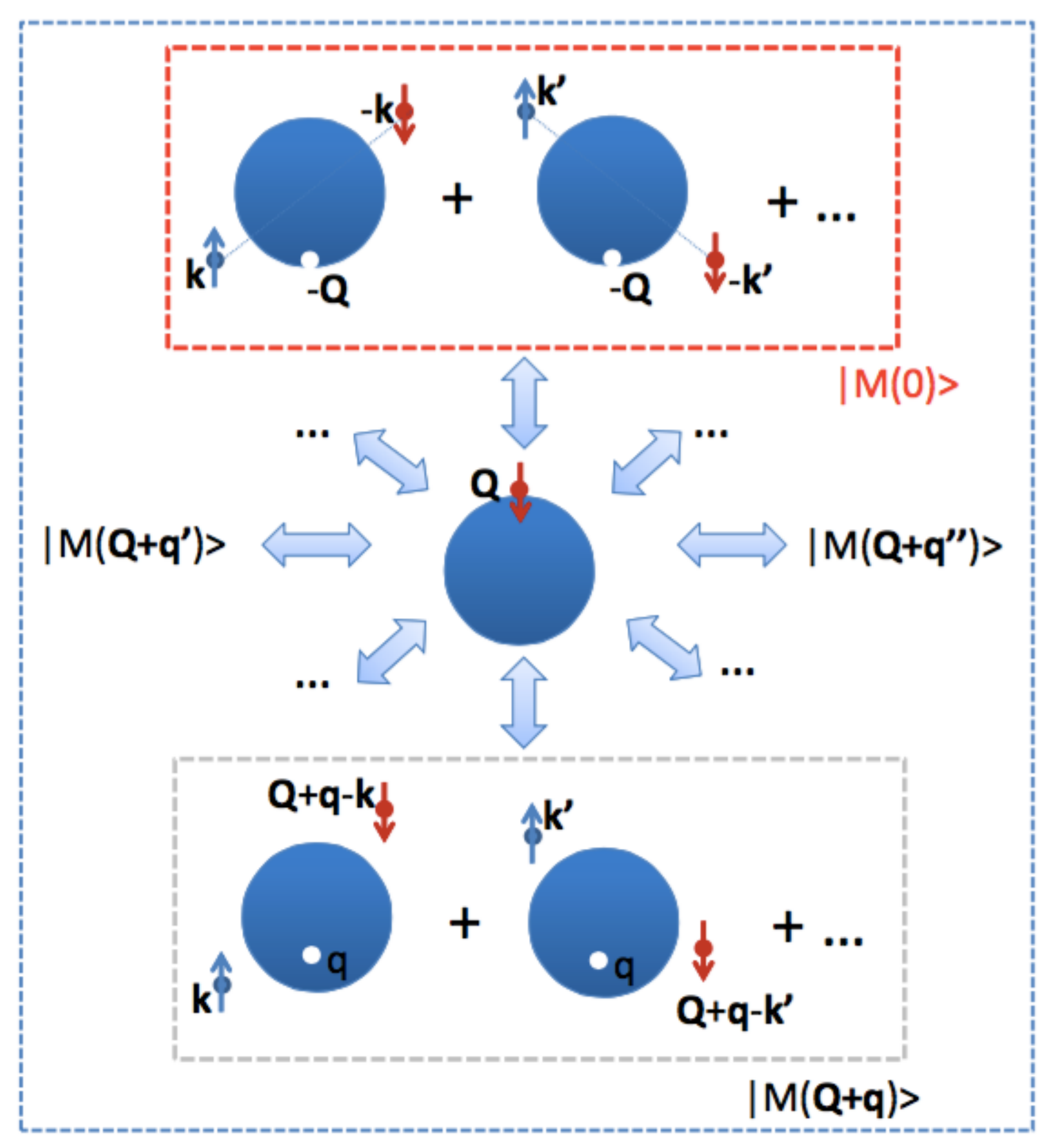}
\caption{(Color Online). Schematics for the instability of molecule state $|M(0)\rangle$ (upper red rectangle), which is constructed by a singlet pair ($\uparrow,\downarrow$) outside the Fermi sea with zero center-of-mass momentum and a hole at $-{\cp Q}=-k_F{\bf e}_z$. $|M(0)\rangle$ can become unstable through coupling to the unperturbed bare state (central picture), and then from this state to other particle-hole excitations with the hole at other momenta, see, for instance $|M({\cp Q}+{\cp q})\rangle$ in the lower gray rectangle. All these configurations compose $|P({\cp Q})\rangle$ state.}  \label{fig_schematic}
\end{figure}

{\it Molecule v.s. finite-$Q$ polaron.} 
By directly comparing (\ref{wf_p})  and (\ref{wf_m}), one can see that if set $Q=k_F$, $\phi_0=0$ and $\phi_{{\cp k},{\cp q}}=\delta_{{\cp q}, -{\cp Q}}\psi_{\cp k}$, then (\ref{wf_p})  exactly reduces to (\ref{wf_m}). In other words, $|M(0)\rangle$ corresponds to only considering a particular type of particle-hole excitations in $|P({\cp Q})\rangle$ with $|{\cp Q}|\equiv Q=k_F$, which will be denoted as $|P(k_F)\rangle$ for short. In such particular excitation, the hole sits right at Fermi surface and points opposite to ${\cp Q}$. However, this type of excitation is not self-closed even in the lowest-order excitation subspace. As shown in Fig.\ref{fig_schematic}, it can scatter back to the bare impurity at ${\cp Q}$ together with an unperturbed Fermi sea, and then couple to other excitations  with holes covering all other momenta inside the Fermi sea\cite{footnote_ex}. 
Here we denote $|M({\cp Q}+{\cp q})\rangle$ as a set of excitation states with total momentum ${\cp Q}$ and the hole sitting at ${\cp q}$. $|P(k_F)\rangle$ is thus a superposition of bare term and all $|M({\cp Q}+{\cp q})\rangle$ states, and it is self-closed up to the lowest particle-hole excitation.  As the coupling between the bare term and different $|M({\cp Q}+{\cp q})\rangle$ will generate a lower variational energy, $|P(k_F)\rangle$ is always energetically more favorable than $|M(0)\rangle$. Therefore, the necessity to introduce such a $|M(0)\rangle$ state seems quite questionable.

\begin{figure}[h]
\includegraphics[width=8.8cm]{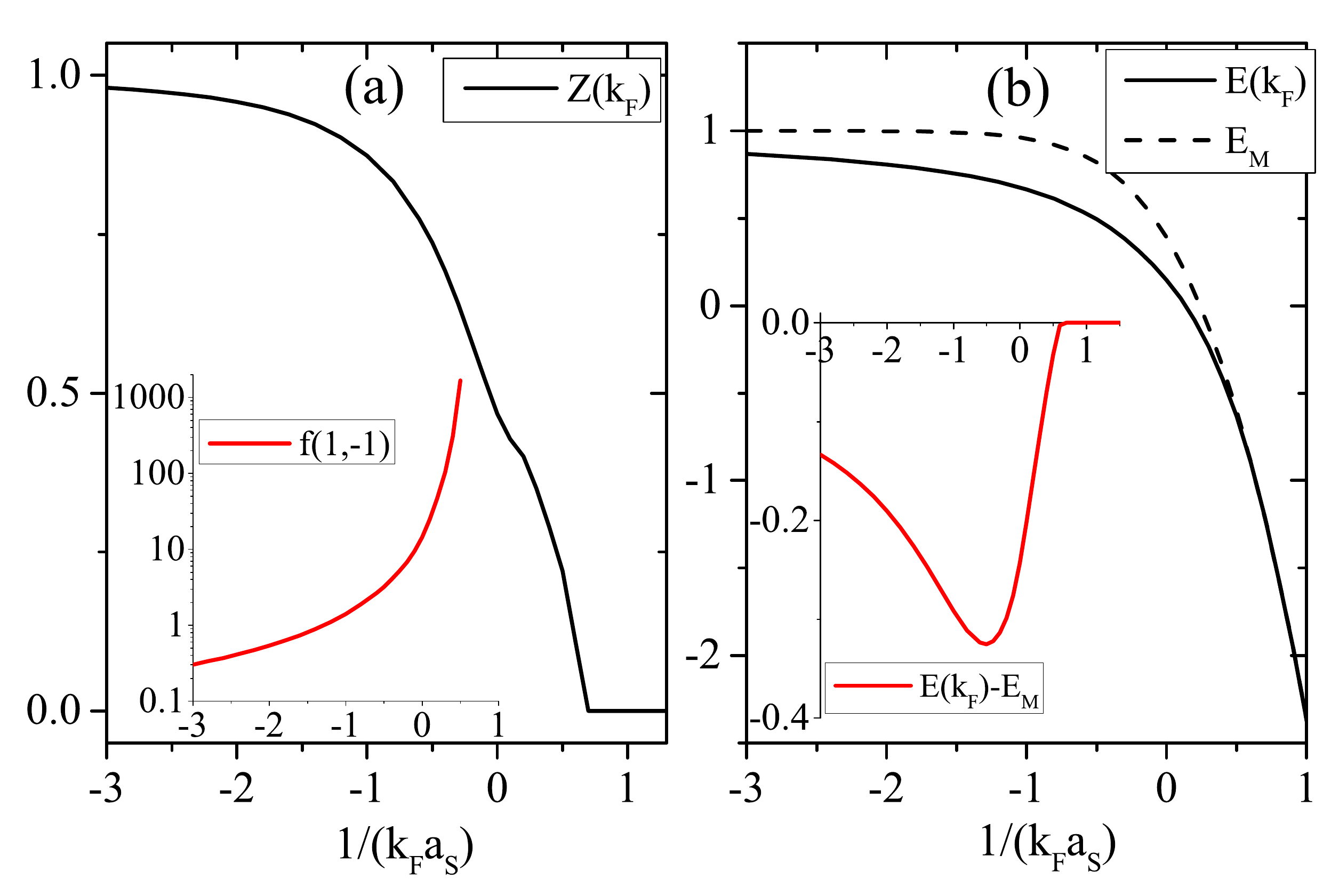}
\caption{(Color Online). (a): residue of $|P(k_F)\rangle$ as a function of $1/(k_Fa_s)$. Inset of (a) shows the weight between molecular component and bare one (see text). (b) Energies of $|P(k_F)\rangle$ ($E(k_F)$) and $|M(0)\rangle$ ($E_M$) as functions of $1/(k_Fa_s)$. Inset shows their difference. The unit of energy is $E_F$.}  \label{fig_Q}
\end{figure}

Here we show that the significance of $|M(0)\rangle $ lies in that it represents the asymptotic limit of $|P(k_F)\rangle$ in the strong coupling regime. To see this, we study the quasi-particle residue of $|P(k_F)\rangle$, $Z(k_F)$, which follows
\begin{eqnarray}
&&Z(k_F)^{-1}=1+\sum'_{{\cp k}{\cp q}}|\phi_{{\cp k}{\cp q}}|^2/|\phi_0|^2 \nonumber\\
&&\ \ \ \ \ =1+\int_0^1 d(q/k_F) \int_{-1}^{1} d(\cos\theta_{\cp q})f(q/k_F, \cos\theta_{\cp q}). \label{Z_F}
\end{eqnarray}
As shown in Fig.\ref{fig_Q}(a), $Z(k_F)$ continuously decreases from unity to nearly zero as the interaction strength increases from weak coupling to $1/(k_Fa_s)\sim 0.8$. This behavior can be traced back to the rapid increase of molecular weight compared to the bare one in $|P(k_F)\rangle$, as given by $f(1,-1)$ in Eq.\ref{Z_F} and shown in the inset of Fig.\ref{fig_Q}(a).  
This indicates a smooth crossover in $Q=k_F$ sector from a polaronic  to a molecular state as the attraction increases. Such crossover is further confirmed by examining the energies of $|P(k_F)\rangle$ and $|M(0)\rangle$, denoted respectively as $E(k_F)$ and $E_M$, as shown in Fig.\ref{fig_Q}(b). Despite a clear deviation at the weak and intermediate couplings, the two energies get closer when the interaction is tuned across resonance, and finally merge together for $1/(k_Fa_s)\gtrsim0.8$. In this strong coupling regime, 
$|M(0)\rangle$ can be justified as a good approximation for $|P(k_F)\rangle$. 

A physical picture to understand above results is as follows. As increasing the attraction strength, the particle-hole excitations become more dominated in $|P(k_F)\rangle$, leading to a reduced $Z(k_F)$. Due to the reduced weight of bare term, the coupling effect between $|M(0)\rangle$ and the other states, as depicted in Fig.\ref{fig_schematic}, becomes less significant. In the strong coupling regime with  $Z(k_F)\sim 0$, the coupling effect is negligible, and $|M(0)\rangle$ is nearly isolated from all other hole excitations (these hole states are all energetically unfavorable as compared to $|M(0)\rangle$). In this limit, $|M(0)\rangle$ can well approximate  $|P(k_F)\rangle$ and $E_M$ becomes identical to $E(k_F)$. We find these results are robust against higher-order particle-hole excitations\cite{future_work}. 

We note that the competition between molecule and finite-$Q$ polaron was also discussed in Ref.\cite{Castin, Parish4}, and their resemblance in strong coupling limit was pointed out in the multi-channel alkali-earth fermions\cite{ZhangWei}. Nevertheless, in these studies the two states were treated separately from independent ansatz, and their generic relation (as depicted in Fig.\ref{fig_schematic}) has not been revealed.  

\begin{figure}[t]
\includegraphics[width=8.8cm]{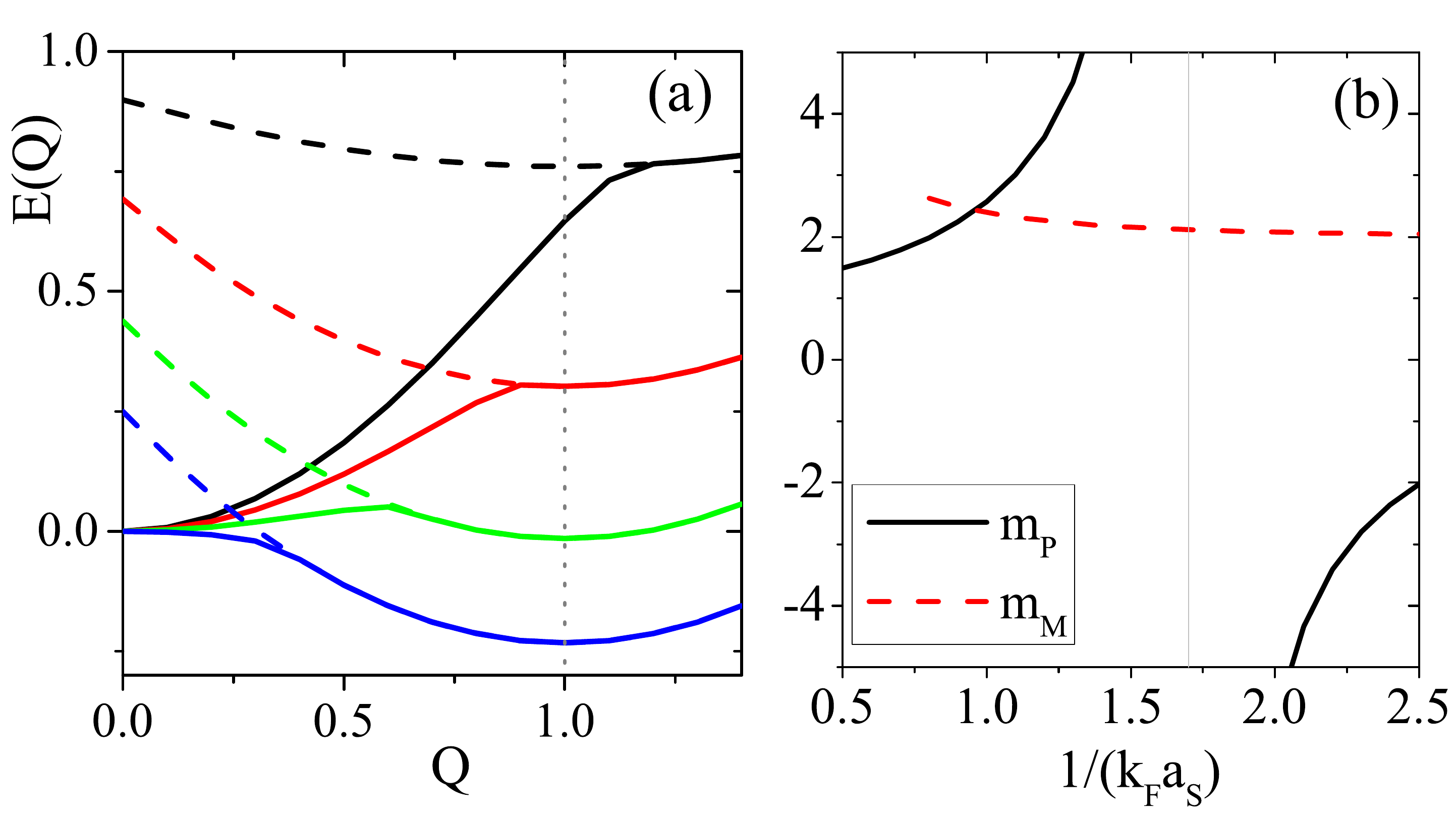}
\caption{(Color Online). First-order transition between $|P(0)\rangle$ and $|P(k_F)\rangle$. (a) Solid lines: energies of $|P(Q)\rangle$ as functions of $Q$ for $1/(k_Fa_s)=0.3,\ 0.8,\ 1.3,\ 2$ (from top to bottom). Dashed lines show results by only including the ${\bf q}=-k_F{\cp e}_z$ hole states in $|P(Q)\rangle$. (b) Effective masses near $Q=0$ and $Q=k_F$, denoted respectively by $m_P$ and $m_M$, as functions of $1/(k_Fa_s)$. The units of momentum and energy are respectively $k_F$ and $E_F$.}  \label{fig_transition}
\end{figure}

{\it Polaron-molecule transition.} Having demonstrated the molecule as an asymptotic limit of finite-$Q$ state, now we are ready to search for possible transition under $|P({\cp Q})\rangle$ throughout all $Q$. In Fig.\ref{fig_transition}(a), we show the dispersion $E(Q)$ for various coupling strengths. One can see that in weak couplings, the only minimum of $E(Q)$ is at $Q=0$, thus $|P(0)\rangle$ is the unique ground state; while increasing $1/(k_Fa_s)$  to $\sim 0.8$, another local minimum appears at $Q=k_F$ but with a higher energy than $E(Q=0)$. At $1/(k_Fa_s)\sim 1.27$, the two minima have the same energy, setting the location of the first-order transition between $|P(0)\rangle$ and $|P(k_F)\rangle$. 
At this point, $|M(0)\rangle$ is already a good approximation for $P(k_F)$. Further increasing $1/(k_Fa_s)$ to $\sim1.7$, $|P(0)\rangle$ becomes a local maximum in the dispersion, and accordingly its effective mass undergoes a resonance from large positive to large negative(see Fig.\ref{fig_transition}(b)). Beyond this point, the only energy minimum occurs at  $Q=k_F$, and the ground state is well approximated by $|M(0)\rangle$. 

Fig.\ref{fig_transition} delivers two important points. First, the literally called polaron-molecule transition for the single impurity system does exist. Nevertheless, the transition by its nature is between impurity systems with different momenta($Q=0$ and $Q=k_F$), rather than between different forms of preset ansatz. Since the transition is between different-$Q$ states, it cannot be accomplished by re-shuffling the Fermi sea via the interaction-induced particle-hole excitations, which conserve the total momentum.  This naturally resolves the theoretical debate on the existence of such transition as in Ref.\cite{Edwards}. 
Second, within an interaction window near their transition, the two-$Q$ states are both locally stable against any momentum fluctuation. This provides the underlying mechanism for their coexistence in realistic systems, as discussed below.

{\it Polaron-molecule coexistence and smooth transition in realistic systems.} 
We consider a small impurity concentration $n_{\downarrow}=0.05n_{\uparrow}$ and a low temperature $T=0.02E_F$ to mimic the realistic condition in experiment. The effects of finite $n_{\downarrow}$ and finite $T$ to the spectroscopy of Fermi polarons were studied in \cite{HuiHu,Levinsen,Tajima1, Tajima2}. Here, to highlight the essential physics of polaron-molecule coexistence, 
we just focus on two possible configurations for the dressed impurities: one is nearby zero-momentum polaron with dispersion $\epsilon^P_{\cp Q}=E_P+{\cp Q}^2/(2m_P)$ ($E_P=E(0)$), which obeys fermionic statistics; the other is nearby $Q=k_F$  with dispersion $\epsilon^M_{\cp Q}=E_M+(|{\cp Q}|-k_F)^2/(2m_M)$, which obeys bosonic statistics and holds for  $1/(k_Fa_s)\geqslant0.8$ when the molecule solution is approached.   Here $m_P$ and $m_M$ are respectively the effective masses of polaron and molecule, as shown in Fig.\ref{fig_transition}(c). These two configurations can stay in equilibrium with each other under the same impurity chemical potential $\mu$, which leads to the  number equation $N_{\downarrow}=N_P+N_M$ with:
\begin{equation}
N_P=\sum_{\cp Q} f_+(\epsilon^P_{\cp Q}-\mu);\ \ N_M=\sum_{\cp Q}f_-(\epsilon^M_{\cp Q}-\mu);\label{N}
\end{equation}
here $f_{\pm}(\epsilon)=(e^{\epsilon/T}\pm 1)^{-1}$. The total energy is
\begin{equation}
E=\sum_{\cp Q} \left(\epsilon^P_{\cp Q}f_+(\epsilon^P_{\cp Q}-\mu)+\epsilon^M_{\cp Q}f_-(\epsilon^M_{\cp Q}-\mu)\right).\label{E}
\end{equation}
In the coexistence region $1/(k_Fa_s)\in(0.8,1.7)$ where both polaron and molecule are locally stable, one can obtain $\mu$ from (\ref{N}) and further $E$ from (\ref{E}), by employing the data of $E_P,\ E_M, m_P, m_M$ as presented in Fig.\ref{fig_transition}. Near the boundaries of coexistence region, Eq.(\ref{N}) automatically guarantees a negligible occupation either on polaron ($N_P\sim 0$) or on molecule ($N_M\sim 0$), due to their visible energy difference $|E_P-E_M|$. Therefore (\ref{N},\ref{E}) can be continuously connected to non-coexistence region, where the system is solely composed by polarons or molecules.

\begin{figure}[t]
\includegraphics[width=8.8cm]{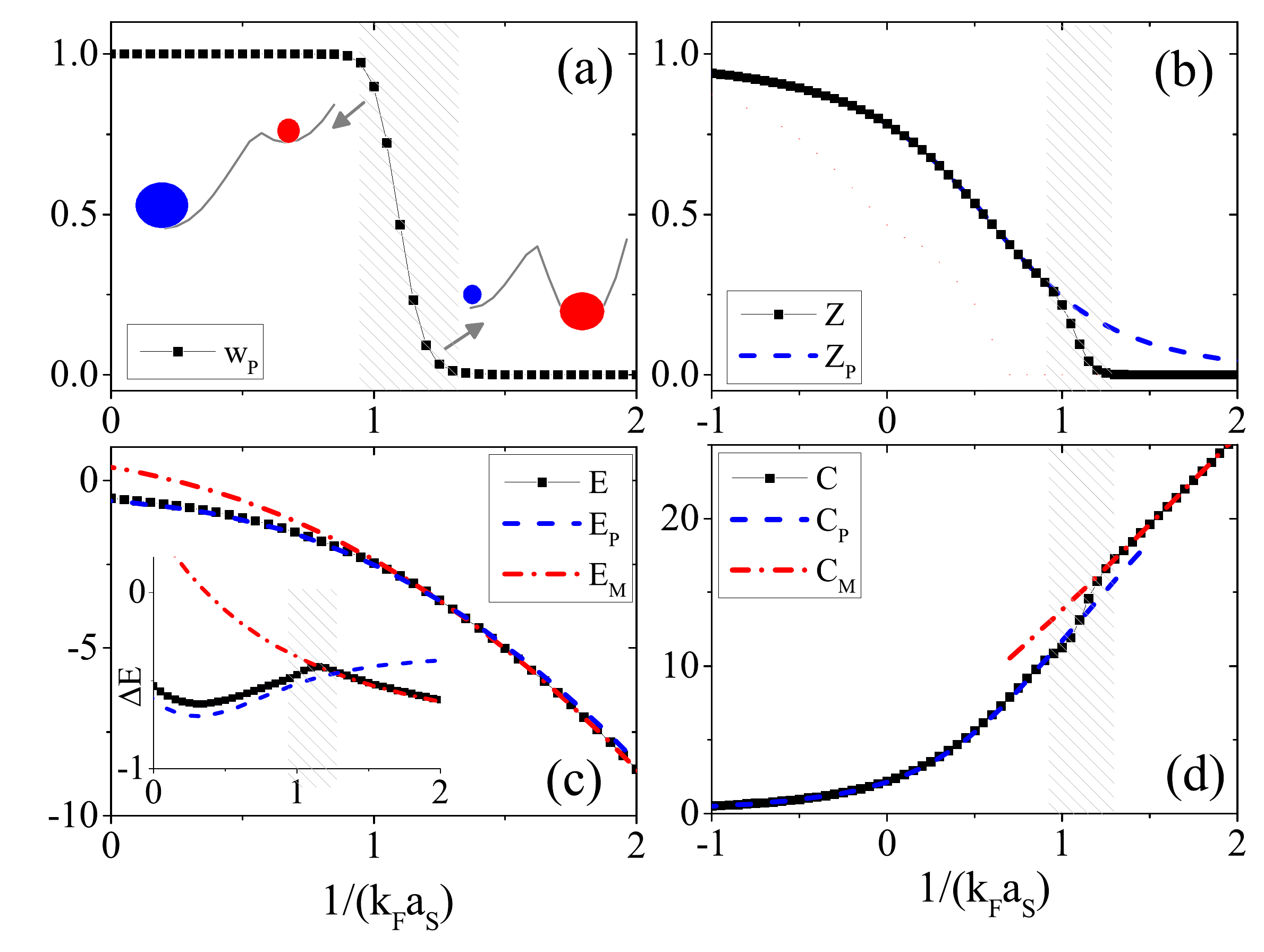}
\caption{(Color Online). Smooth polaron-molecule transition with a finite impurity density $n_{\downarrow}=0.05n_{\uparrow}$ and at temperature $T=0.02E_F$. (a,b,c,d) respectively show the polaron weight $w_P$, residue $Z$, energy $E$ scaled by $N_{\downarrow}E_F$,  and contact $C$ scaled by $(2N_{\downarrow}k_F)$.  For comparison, blue dashed (red dashed-dot) lines show results based on polaron (molecule) ansatz. The shaded areas mark the region of  visible polaron-molecule coexistence. Inset of (a) shows two typical situations when the polaron (left blue) or molecule (right red) dominates in the impurity occupation, depending on their energy comparison. Inset of (c) shows the energy per impurity shifted by two-body binding energy $E_{2b}=-1/(ma_s^2)$, for a better view of smooth transition. 
}  \label{fig_crossover}
\end{figure}

In Fig.\ref{fig_crossover}(a-d), we show the polaron weight $w_P\equiv N_P/N_{\downarrow}$, residue $Z=Z_P w_P$ ($Z_P=Z(0)$), energy $E$, and contact $C=(4\pi m)dE/d(1/a_s)$. We can see that all quantities evolve continuously from the weak to strong coupling regime, consistent with experimental observations\cite{Sagi}. In the weak (strong) coupling regime, both $E$ and $C$ are well explained by the polaron (molecule) results (see dashed lines). All these features demonstrate a smooth polaron to molecule transition in realistic impurity systems. 
In particular, we note that $Z_P$ shows an obvious reduction from unity to zero within $1/(k_Fa_s)\in[0.9,1.3]$, as marked by shaded area in Fig.\ref{fig_crossover}, which sets the region for visible polaron-molecule coexistence. Such coexistence washes out all the discontinuities at the first-order transition, and turns it to a smooth one in realistic system. We also note that the visible coexistence terminates at $1/(k_Fa_s)\sim 1.3$, very close to the transition point $\sim 1.27$. It can be attributed to the bosonic enhancement, where particles tend to condense as molecular bosons once across the transition.  This implies the experimentally measured zero-crossing location of $Z$\cite{Zwierlein, Sagi} are indeed close to the transition point.

\textit{Discussion and outlook.}
Our results can be further improved by including the second-order particle-hole excitation in variational ansatz, which has been studied in $Q=0$ sector and shown to be as accurate as Monte Carlo method\cite{Combescot2, Leyronas,Prokofev}. Our preliminary study on finite-$Q$ sectors confirms that their inclusion does not change qualitatively the essential physics revealed in this work, except for shifting the polaron-molecule transition point and coexistence region to weaker couplings\cite{future_work}. Moreover, it is noted that the trap inhomogeneity in existing experiments\cite{Sagi, Zwierlein}  can also contribute to smoothening the transition, as polarons and molecules can appear in different locations inside the trap even without the mechanism discussed in this work.  Therefore, a more transparent testbed for our theory is a homogeneous Fermi gas, which has become accessible in cold atoms laboratories\cite{Zwierlein2,Moritz}.

Finally, our analyses on the molecule instability, the competition between different-$Q$, and the smoothening mechanism can in principle be applied to Fermi polarons in low dimensions. In 2D, the existence of polaron-molecule transition for single impurity system is still an open question  given different conclusions from variational methods\cite{Parish, Parish2, Pethick, ZhangWei2} and quantum Monta Carlo\cite{MC_2d_1, MC_2d_2, MC_2d_3}. 
In 1D, the situation is even more intriguing due to contributions from hole scattering\cite{Combescot3}. Exact solutions have shown a smooth crossover instead of a sharp transition, and the effective mass never displays a resonance\cite{McGuire, Guan}. In this context, the physics of competition/conversion between polaron and molecule in reduced dimensions still requires a careful study in future.

{\it Acknowledgements.} The work is supported by the National Key Research and Development Program of China (2018YFA0307600, 2016YFA0300603), the National Natural Science Foundation of China (11534014, 12074419), and the Strategic Priority Research Program of Chinese Academy of Sciences (XDB33000000).

{\bf Note added:} During the submission of our paper, the updated experiment by Israel group\cite{Sagi} has newly adopted a theory similar to ours to explain the observed smooth polaron-molecule transition. The idea of the polaron-molecule coexistence and equilibrium in the presence of a finite impurity density and at finite temperature is consistent with ours. However, in Ref.\cite{Sagi} the polaron and molecule dispersions  are both centered at zero momentum while here their centers are shown to be differed by $k_F$.


\begin{thebibliography}{99}
\bibitem{footnote_MC}Even for the diagrammatic Monte-Carlo studies\cite{Prokofev}, the polaron and molecule were treated separately and their energies were individually extracted from different quantities, namely, the single-particle self energy and two-particle scattering matrix.
\end{thebibliography}
\end{document}